\documentclass[%
aip,
amsmath,amssymb,
reprint,%
]{revtex4-1}

\usepackage{graphicx}
\usepackage{dcolumn}
\usepackage{bm}

\usepackage[utf8]{inputenc}
\usepackage[T1]{fontenc}
\usepackage{mathptmx}
\usepackage{etoolbox}

\makeatletter
\def\@email#1#2{%
\endgroup
\patchcmd{\titleblock@produce}
{\frontmatter@RRAPformat}
{\frontmatter@RRAPformat{\produce@RRAP{*#1\href{mailto:#2}{#2}}}\frontmatter@RRAPformat}
{}{}
}%
\makeatother
\begin{document}

\preprint{AIP/123-QED}

\title[Sample title]{NiCrAl piston-cylinder cell for magnetic susceptibility measurements under high pressures in pulsed high magnetic fields}
\author{Katsuki Nihongi}
\affiliation{Center for Advanced High Magnetic Field Science (AHMF), Graduate School of Science, Osaka University, Toyonaka, Osaka 560-0043, Japan}%

\author{Takanori Kida}%
\affiliation{Center for Advanced High Magnetic Field Science (AHMF), Graduate School of Science, Osaka University, Toyonaka, Osaka 560-0043, Japan}%

\author{Yasuo Narumi}
\affiliation{Center for Advanced High Magnetic Field Science (AHMF), Graduate School of Science, Osaka University, Toyonaka, Osaka 560-0043, Japan}%

\author{Nobuyuki Kurita}
\affiliation{Department of Physics, Tokyo Institute of Technology, Meguro-ku, Tokyo 152-8551, Japan}%

\author{Hidekazu Tanaka}
\affiliation{Innovator and Inventor Development Platform (IIDP) Tokyo Institute of Technology Nagatsuda, Midori-ku, Yokohama 226-8502, Japan}%

\author{Yoshiya Uwatoko}
\affiliation{The Institute for Solid State Physics (ISSP), The University of Tokyo, Kashiwa, Chiba 277-8581, Japan}%

\author{Koichi Kindo}
\affiliation{The Institute for Solid State Physics (ISSP), The University of Tokyo, Kashiwa, Chiba 277-8581, Japan}%

\author{Masayuki Hagiwara}
\affiliation{Center for Advanced High Magnetic Field Science (AHMF), Graduate School of Science, Osaka University, Toyonaka, Osaka 560-0043, Japan}%
\email{hagiwara@ahmf.sci.osaka-u.ac.jp}

\date{\today}%

\begin{abstract}

We developed a metallic pressure cell made of nickel-chromium-aluminum (NiCrAl) for use with a non-destructive pulse magnet and a magnetic susceptibility measurement apparatus with a proximity detector oscillator (PDO) in pulsed magnetic fields of up to 51 T under pressures of up to 2.1 GPa. Both the sample and sensor coil of the PDO were placed in the cell so that the magnetic signal from NiCrAl would not overlay the intrinsic magnetic susceptibility of the sample. A systematic investigation of the Joule heating originating from metallic parts of the pressure cell revealed that the temperature at the sample position remains at almost 1.4 K until approximately 80 $\%$ of the maximum applied magnetic field ($H_{\rm max}$) in the field-ascending process (e.g., 40 T for $H_{\rm max}$ of 51 T). The effectiveness of our apparatus was demonstrated, by investigating the pressure dependence of the magnetization process of the triangular-lattice antiferromagnet Ba$_3$CoSb$_2$O$_9$.

\end{abstract}

\maketitle

\section{Introduction}

Extreme conditions, such as high pressure, high magnetic field, and low temperature, are occasionally required to search for new properties and phenomena in condensed-matter materials. For instance, the ground states of geometrically frustrated magnets (GFMs) infinitely degenerate at low temperatures, and exotic physical phenomena such as a quantum spin-liquid state, and quantum phase transitions have been reported under extreme conditions\cite{Anderson.spinliquid, Balents.spinliquid, Nikuni}. In GFMs, a high magnetic field lifts the degeneracy and sometimes induces exotic magnetic phases. High pressure alters the magnetic anisotropy and exchange interactions between magnetic ions in a magnetic material by shrinking its crystal lattice. Recently, the triangular-lattice antiferromagnet Cs$_2$CuCl$_4$, one of the GFMs, was reported to exhibit multiple magnetic-field-induced phase transitions under high pressure at low temperatures\cite{Zvyagin}. Therefore, experimental techniques that can be used under these extreme conditions are desirable to clarify the physical properties of condensed-matter materials.

The development of measurement techniques under multiple extreme conditions has been undertaken at pulsed high magnetic field facilities. Thus far, the magnetization curves of several magnetic materials measured by a conventional induction method using pick-up coils were reported under pressures of up to 0.95 GPa in pulsed magnetic fields of up to 50 T\cite{hamamoto2000metamagnetic, tahara2020vanishment, matsunaga2010high}. In these studies, a non-destructive pulse magnet and a self-clamped piston-cylinder cell (PCC) made of beryllium-copper (CuBe) or nickel-chromium-aluminum (NiCrAl) were utilized. The magnetization signal was detected by winding pick-up coils with approximately100 turns around the exterior of the PCC (Fig.\ref{fig1}(c)). Therefore, the measurement signals were degraded by the low sample filling rate in the pick-up coils and the noise induced by the eddy current in the metallic parts of PCC caused by pulsed magnetic fields. Moreover, the eddy current causes Joule heating, resulting in the temperature rise of the sample. Hamamoto et al. reported the effect of pressure on the metamagnetic transition in CeRh$_2$Si$_2$ above 6 K in pulsed high magnetic fields using a CuBe PCC\cite{hamamoto2000metamagnetic}. The temperature dependence of the metamagnetic transition field on CeRh$_2$Si$_2$ was reported to be almost independent of the temperature, at least below 15 K, but the temperature change of the sample during the magnetic-field sweep was unknown. In magnetic materials such as GFMs with a low N\'{e}el temperature $T_{\rm N}$, the magnetic properties are often sensitive to temperature changes at low temperatures and the measurements to determine these properties need to be taken below the temperature of liquid helium ($\sim$ 4.2 K).  However, it is difficult to use the aforementioned apparatus to study GFMs.

To suppress the Joule heating, the cell body of the PCC was made of NiCrAl alloy with a lower conductivity than the CuBe alloy. In addition, the tensile strength of the NiCrAl alloy ($\sim$ 2.37 GPa at room temperature (RT)) is higher than that of the CuBe alloy ($\sim$ 1.35 GPa at RT)\cite{walker1999nonmagnetic}. However, the magnetic susceptibility of the NiCrAl alloy was approximately ten times larger than that of the CuBe alloy\cite{mori2007kouatsugijutsuhandbook}. Therefore, the practical use of a NiCrAl PCC is limited to materials with large magnetization magnitudes. To overcome these problems, we developed magnetometry based on a radio frequency (RF) technique using a proximity detector oscillator (PDO) \cite{MMAltarawneh, Ghannadzadeh}.

The PDO is an inductance ($L$)-capacitance ($C$) self-resonating LC tank circuit based on the widely available proximity detector chip used in modern metal detectors. This device can detect the magnetic susceptibility and/or electrical conductivity of a sample in pulsed high magnetic fields\cite{MMAltarawneh, Ghannadzadeh}. In this technique, the inductance change of a small sensor coil with tens of turns in the LC tank circuit is measured when a magnetic field is applied. The resonance frequency of the LC tank circuit at zero field was $f_0 =1/2\pi \sqrt{LC}$. When a sample is placed in the sensor coil, $L$ changes depending on its magnetic susceptibility and/or electrical conductivity of the sample in the magnetic field. Hereafter, we call this technique as LC method. The LC method detects the change in the resonance frequency ($\Delta f$) corresponding to the change in $L$. When the sample is a magnetic insulator, $\Delta f$ is proportional to the change in the dynamic magnetic susceptibility ($\chi = \Delta M/ \Delta H$), as follows:

\begin{equation}
\frac{\Delta f}{ f_0}=-\frac{\Delta L}{2L}\propto - \frac{1}{2} \frac{V_{\rm s}}{V_{\rm c}} 4 \pi \chi,
\label{eq1}
\end{equation}

\noindent
where $V_{\rm s}$ is the volume of the sample inside the sensor coil, and $V_{\rm c}$ is the inside volume of the sensor coil. According to Eq. \ref{eq1}, the absolute value of $\Delta f$ increases as the sample filling rate increases against the sensor coil ($V_{\rm s}/V_{\rm c}$). The sensor coil typically consists of only 5$\sim$30 turns with a diameter as small as 300 $\mu$m. Therefore, an effective approach is to place the small sensor coil, including the sample, inside the small interior space of a high-pressure cell, because the sensor coil does not detect the magnetization of the pressure cell.

Magnetic susceptibility measurements, conducted under high pressure by utilizing the LC method in static magnetic fields, have been reported \cite{Zvyagin,wehinger2018giant,shi2022discovery}. However, such measurements in pulsed magnetic fields were rarely reported. Recently, Sun et al. developed a diamond anvil cell (DAC) fabricated mainly of insulating composites that minimize Joule heating in pulsed high magnetic fields. They performed magnetic susceptibility measurements of the quantum antiferromagnet $\rm{[Ni(HF_2)(pyz)_2]SbF_6}$ in pulsed magnetic fields of up to 65 T under pressure of up to 5.4 GPa by the LC method\cite{sun2020high}. Because of the small sample space in this pressure cell (less than 0.01 mm$^3$), the sensor coil was limited to a diameter of 150 $\mu$m and a maximum of four turns, and the sample size was too small, complicating attempts to increase the sensitivity of the measurement by increasing the number of turns.

In this study, we designed a NiCrAl PCC that suppresses the effect of Joule heating on a sample in pulsed high magnetic fields and established a magnetic susceptibility measurement system based on the LC method for use under multiple extreme conditions. Although the PCC generally generates lower pressures than a DAC, the sensitivity of the measurements can be increased by adjusting the number of turns of the coil because of the larger interior space in the PCC. To demonstrate the effectiveness of this apparatus for the study of GFMs, we examined the magnetization processes of the triangular-lattice antiferromagnet Ba$_3$CoSb$_2$O$_9$, a GFM with $T_{\rm N}$ = 3.8 K at 1.4 K. The magnetic susceptibility was measured under high pressure in pulsed high magnetic fields.

\section{Pressure cell design and setup}

\begin{figure}[h]
\includegraphics[keepaspectratio, scale=0.5]{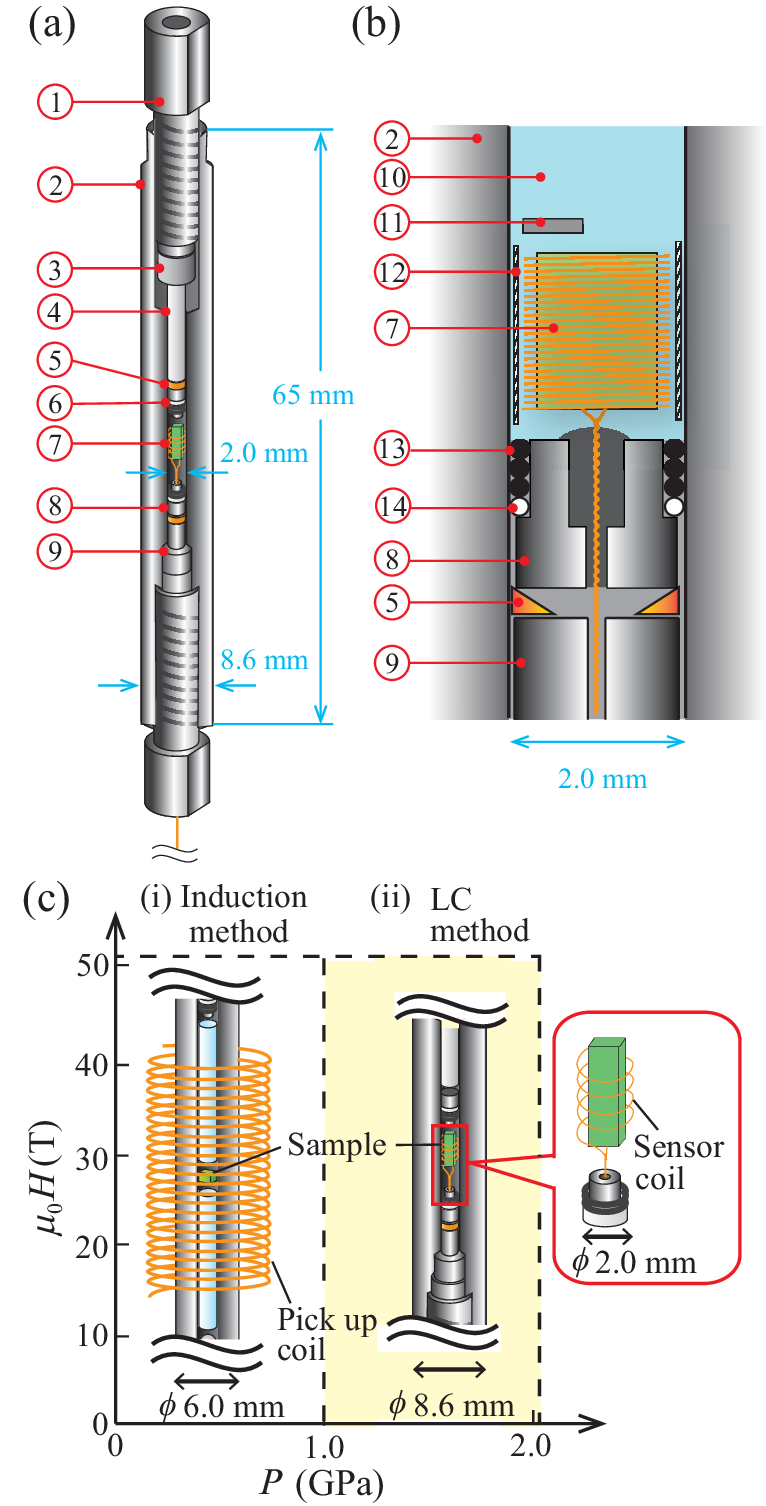}%
\caption{\label{fig1}(a) Schematic view of the piston-cylinder cell (PCC) made of NiCrAl alloy for the magnetic susceptibility measurements in pulsed high magnetic fields under high pressure. (b) Cross-sectional view around the sample space in the PCC. The components are as follows: (1) pressure-clamp bolt ; (2) cylinder with the inner (2.0 mm) and outer (8.6 mm) diameters ; (3) upper piston backup ; (4) ZrO$_2$ piston ; (5) Cu ring ; (6) upper plug with a Teflon ring and three O rings ; (7) a sample (green) around which the sensor coil is directly wound (orange lines) ; (8) lower plug with a stepped hole ; (9) lower piston backup ; (10) pressure medium (Daphne 7373) ; (11) Sn manometer ; (12) Teflon tube ; (13) O-ring ; (14) Teflon ring. (c) Magnetic field and pressure range of magnetic-susceptibility-measurement techniques using the induction and LC methods.}
\end{figure}

Figure \ref{fig1}(a) shows a schematic view of the NiCrAl PCC for the magnetic susceptibility measurements in pulsed high magnetic fields. The cylinder of the PCC, pressure-clamp bolts, plugs, and piston backups were made of NiCrAl alloy. The pressure in the sample space was determined from the pressure dependence of the superconducting transition temperature of Sn \cite{smith1967will}. The pressure cell was inserted into a SQUID magnetometer (Quantum Design, MPMS-XL 7), and the change in the superconducting transition temperature of the Sn manometer was investigated under high pressure. The outer diameter of the cylinder was 8.6 mm, allowing compatibility with the SQUID magnetometer with an inner bore diameter of 9 mm. Moreover, this size was also suitable for insertion into a $^4$He cryostat with an inner bore diameter of 10 mm in a liquid-helium bath. The length of the cylinder was 65 mm; therefore, the length of the sample space was 10 mm under maximum pressure. 

A cross-sectional view of the sample space in the PCC is shown in Fig.\ref{fig1}(b). The pressure medium was Daphne 7373 (Idemitsu Kosan Co., Ltd.). The sample space is filled with Daphne 7373 sealed by NiCrAl plugs with O-rings, Teflon rings, and Cu rings. Cu wires ($\sim$ 100 $\mu$m) pass through the stepped hole of the lower plug filled with STYCAST 2850FT to prevent leaking pressure medium. At RT, the pressure medium remained in the liquid state up to a pressure of approximately 2 GPa. For this pressure medium, the pressure difference between 4.2 and 300 K is reported to be approximately 0.15 GPa, irrespective of the initial pressures at 300 K\cite{murata1997pt}. The sample is usually molded to a height of 5 mm and a diameter of 1.4 mm or less. A Teflon tube with inner and outer diameters of 1.6  and 1.8 mm, respectively, and a length of approximately 10 mm covers the sample and the sensor coil to prevent direct contact between the sample and the inner wall of PCC. The Sn manometer is inserted in the Teflon tube. High pressure was applied to the pressure cell through the piston that was clamped using a pressure clamp bolt at RT. In our preliminary experiments, a NiCrAl PCC with inner and outer diameters of 2.0 and 6.0 mm, respectively, generated pressure of 0.8 GPa for a maximum applied force of nearly 300 kgf. The advantage of this arrangement is that the applied force can be increased by increasing the thickness of the PCC cylinder. In practice, setting the inner diameter to 2.0 mm and expanding the outer diameter to 8.6 mm enabled a maximum applied force of approximately 1000 kgf. Consequently, the NiCrAl PCC has achieved a maximum pressure of $P$ = 2.10 $\pm$ 0.02 GPa.

\begin{figure}[t]
\includegraphics[keepaspectratio, scale=0.65]{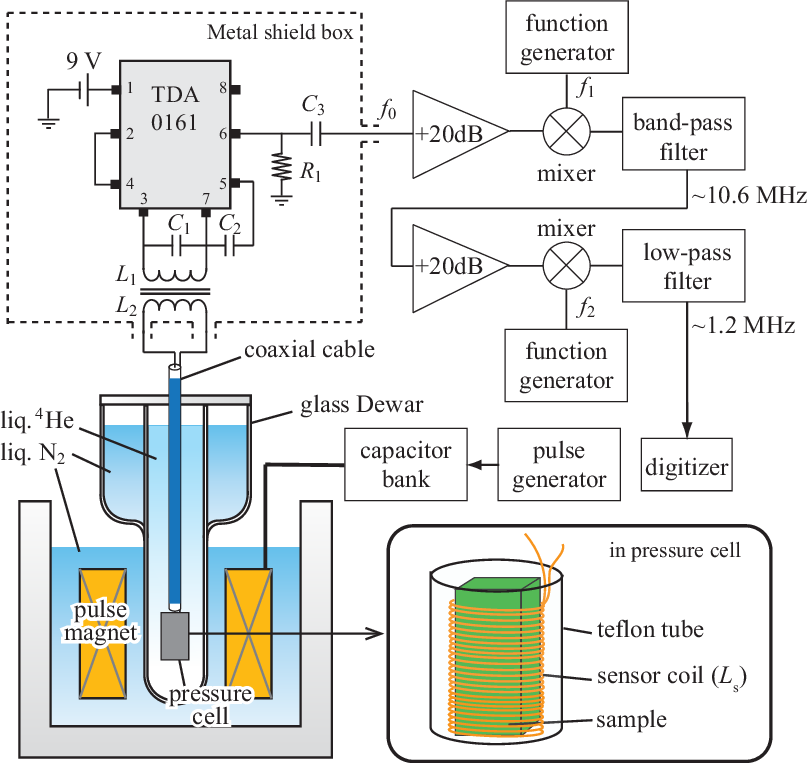}%
\caption{\label{fig2}Block diagram of the LC method for magnetic susceptibility measurements in pulsed magnetic fields under high pressure with the PDO technique. $L_{\rm s}$ is the sensor coil inductance, and is coupled to the proximity detector chip through inductances $L_1$ and $L_2$. $C_1$ and $R_1$ represent capacitors and resistors, respectively. Our previous magnetic susceptibility measurements of Ba$_3$CoSb$_2$O$_9$ in pulsed magnetic fields under high pressure enabled the magnitudes of $L_1$, $L_2$ and $L_{\rm s}$ to be estimated to be 0.2 $\mu$H, 0.2 $\mu$H, and 0.9 $\mu$H, respectively. Additionally, $C_1$, $C_2$, $C_3$ and $R_1$ are 69 pF, 39 pF, 39 pF, and 274 $\Omega$, respectively.}
\end{figure}

Figure \ref{fig2} shows a block diagram of the magnetic susceptibility measurement apparatus for pulsed magnetic fields under high pressure using the PDO. Pulsed magnetic fields were generated using a non-destructive pulse magnet and a capacitor bank installed at the AHMF at Osaka University. The pulse magnet with a bore diameter of 17$\sim$18 mm is immersed in liquid nitrogen to lower the electrical resistance and cool down the magnet after the high-field generation. The pulse magnet was capable of generating pulsed magnetic fields of up to 51 T with a pulse duration of 35 milliseconds (ms). The glass Dewar container consisted of a liquid-helium bath containing the PCC with the sample, a vacuum insulation space, and a liquid nitrogen bath. The sample space can be cooled to 1.4 K at the lowest by evacuating the liquid helium bath with liquid $^4$He.

The design of the PDO circuit surrounding the metal shield box, shown in Fig.\ref{fig2}, was based on designs in previous reports of Refs.\onlinecite{MMAltarawneh, Altarawneh, Ghannadzadeh}. To obtain an intense PDO signal, the sensor coil ($L_{\rm s}$) with 40 $\mu$m diameter Cu wire was directly wound around the sample to get $V_{\rm s}/V_{\rm c} \approx 1$ in Eq.1 and the number of turns was adjusted accordingly. In this study, the sensor coil was wound to $\sim$25 turns for the small sample (typical size is $\sim 1\times1\times 5$ ${\rm mm}^3$) that can be inserted into the PCC. The sensor coil placed in the helium bath was connected to the PDO circuit in the metal shield box at RT with a coaxial cable (Lake Shore Crytronics Inc., Ultra-Miniature Coaxial Cable type C) of approximately 1 m. The resonance frequency in the entire PDO circuit, including the sensor coil and coaxial cable depends on the effective inductance ($L_{\rm eff}$) composed of $L_{\rm s}$, $L_1$ and $L_2$; the mutual inductance $L_{\rm m}$ among the coils; and the connecting coaxial cable ($L_{\rm coax}$). The total effective inductance $L_{\rm eff}$ is given by,

\begin{equation}
L_{\rm eff} = L_1(1- \frac{L_{\rm m}^2}{L_1(L_2+L_{\rm s}+L_{\rm coax})}).
\label{eq2}
\end{equation}
\\
\noindent
In this setup, the resonant frequency in zero field ($f_0$) was 35$\sim$42 MHz. The output signals ($f ({\rm \mu_0}H) = f_0+\Delta f$) measured in pulsed magnetic fields were amplified and sent to two-stage frequency mixing ($f_1$, $f_2$), and were filtered to remove high-frequency components. The frequency of the output signal ($\sim$42 MHz) loaded into the digitizer is down-converted to 1.2 MHz. The signal was stored  in the digitizer at a rate of 50 MS/s (MS: mega-samples), with one wave consisting of approximately 300 data points, which was sufficient to construct the correct waveform. The average frequency at each point for the descrete magnetic field was made from 3$\sim$5 successive waves. Consequently, the actual sampling rate corresponded to approximately 240$\sim$400 kS/s (kS: kilo-samples).

\section{Effect of Joule heating}

\begin{figure}[ht]
\includegraphics[keepaspectratio, scale=0.5]{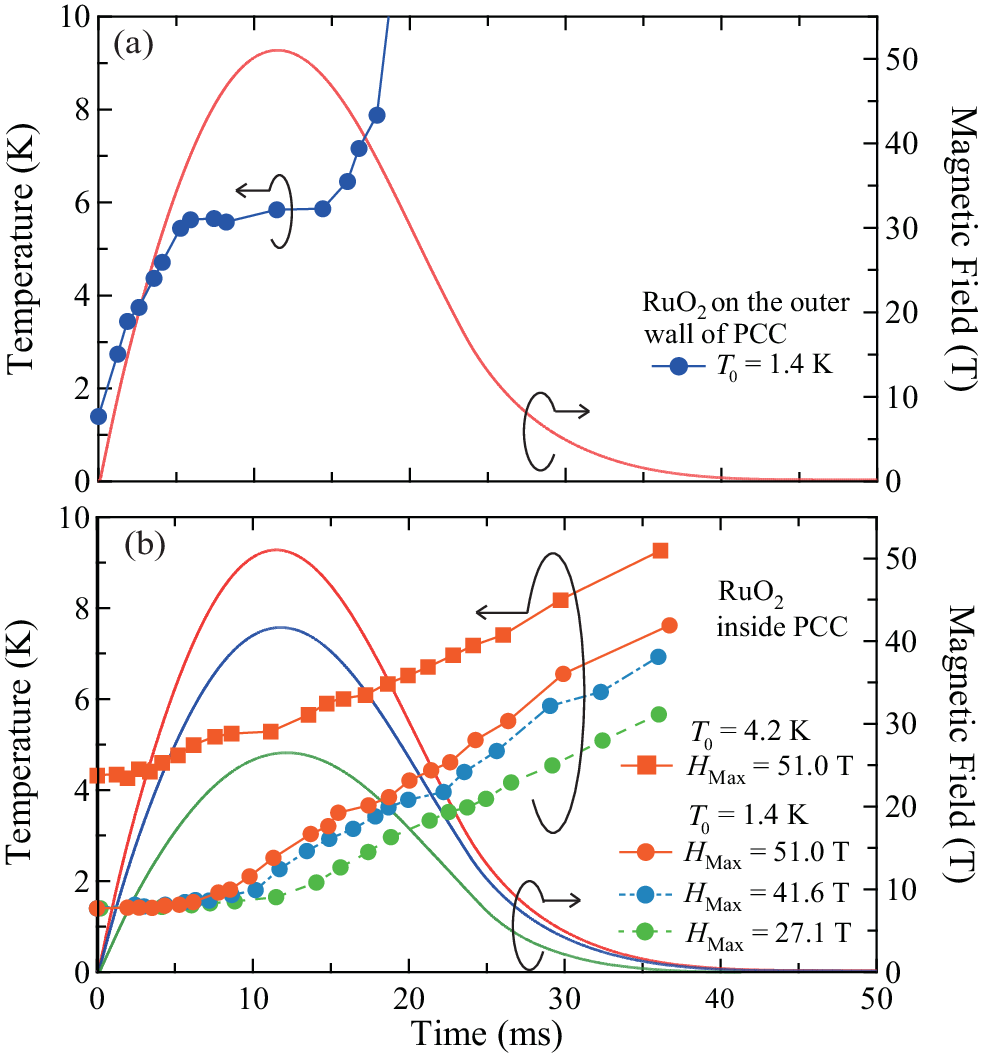}%
\caption{\label{fig3}Temperature changes at the sample position inside the PCC and on the outer wall of the PCC in pulsed magnetic fields. (a) Temperature change on the outer wall of the PCC starting at $T_0$ = 1.4 K and pulsed-magnetic-field profile at the maximum field of 51.0 T. (b) Temperature change at the sample position inside the PCC starting at $T_0$ = 1.4 K and 4.2 K in various maximum fields. The orange solid, light blue dot-dashed, and yellow-green dotted lines represent the temperature changes inside the PCC for the case of $H_{\rm max}$ = 51.0, 41.6, and 27.1 T, respectively.}
\end{figure}

To evaluate the amount of heat transferred from the heated pressure cell to a sample in the presence of high magnetic field, we investigated the temperature change in the sample space in pulsed magnetic fields utilizing a commercially available RuO$_2$-tip resistor (KOA Co. Ltd, typical resistance is 560 $\Omega$ at RT) as a thermometer. The magnetoresistance of this RuO$_2$-tip resistor was calibrated in pulsed magnetic fields below 10 K, and the tip resistor was placed in the sample space filled with Daphne 7373 or on the outer wall of the PCC. The PCC was inserted into the glass Dewar container filled with liquid $^4$He ($\sim$1.4 K) as shown in Fig.\ref{fig2}.

\begin{figure}[b]
\includegraphics[keepaspectratio, scale=0.55]{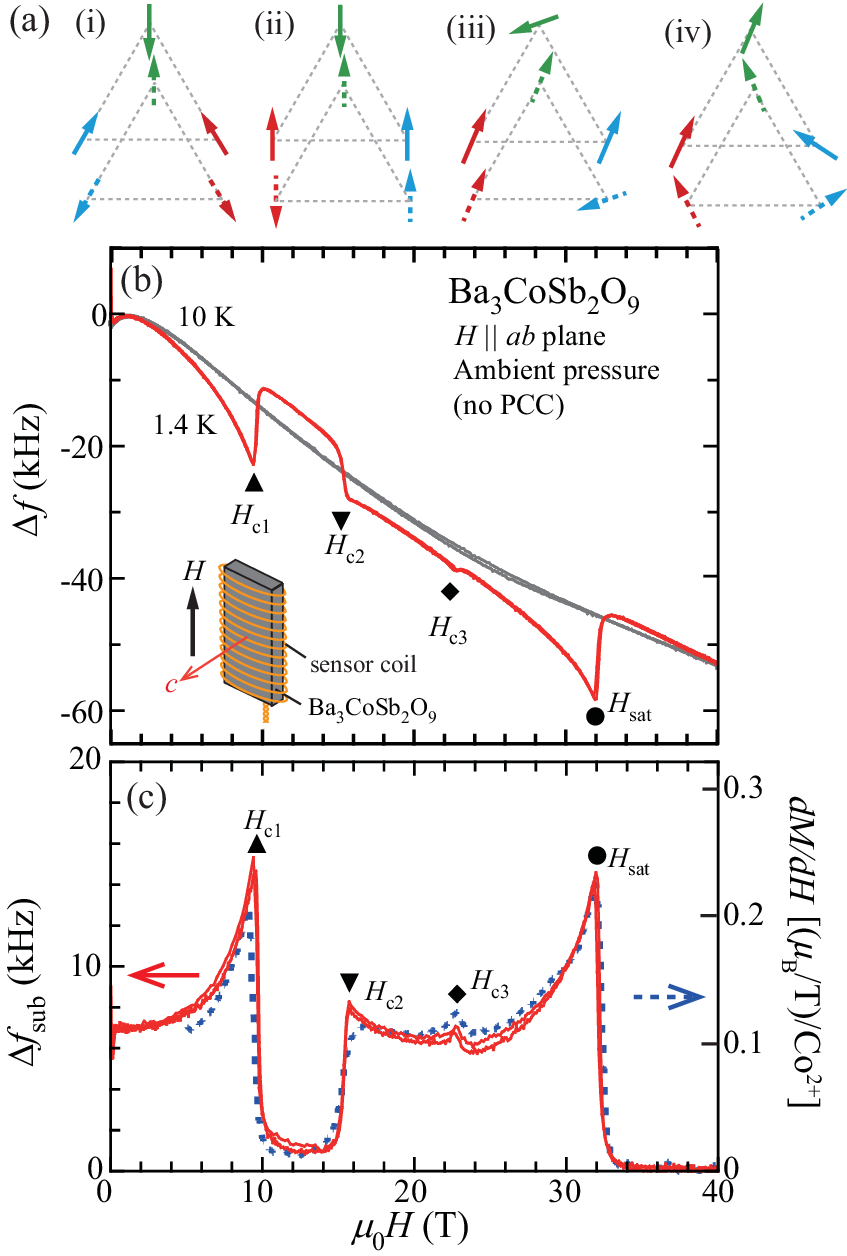}%
\caption{\label{fig4}(a) Magnetic structure of Ba$_3$CoSb$_2$O$_9$ for $H \parallel ab$ plane. Solid arrows and dotted arrows of the same color represent spins at the sublattice vertices in neighboring triangular lattice layers, respectively. (i) Y coplanar state (ii) up-up-down state (iii) V state (iv) V$^{\prime}$ state; (b) the change in the resonance frequencies ($\Delta f$) for $H \parallel ab$ plane of Ba$_3$CoSb$_2$O$_9$ at 1.4 K and 10 K under ambient pressure without the PCC. Inset: schematic view of the sensor coil including Ba$_3$CoSb$_2$O$_9$. (c) $\Delta f_{\rm sub}$-$H$ curve (red solid line) and $dM/dH$ measured with the induction method in Ref.~\onlinecite{susuki2013magnetization} (blue dotted line).}
\end{figure}

\begin{figure*}[t]
\includegraphics[keepaspectratio, scale=0.5]{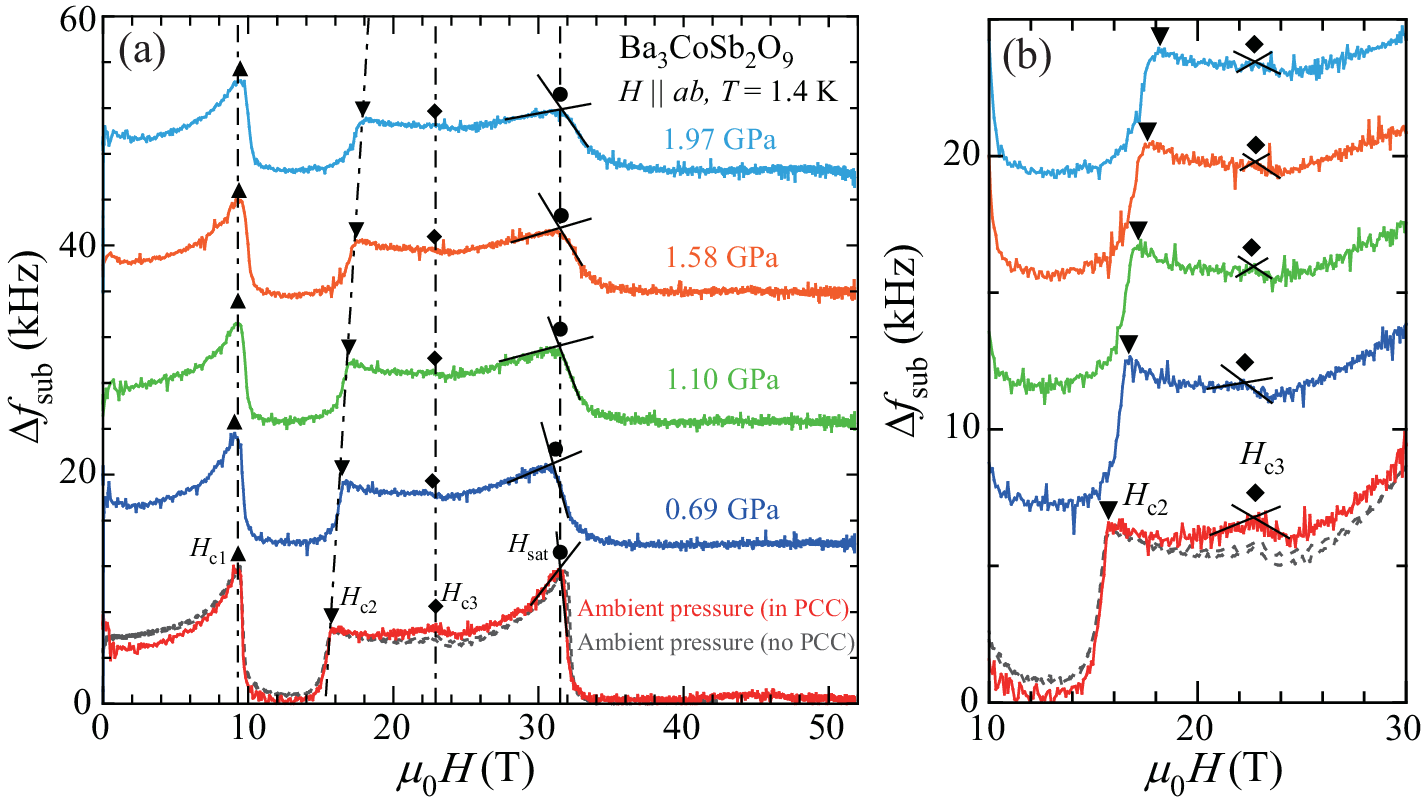}%
\caption{\label{fig5}(a) $\Delta f_{\rm sub}$-$H$ curves for $H \parallel ab$ plane of Ba$_3$CoSb$_2$O$_9$ at 1.4 K at various pressures. The dot-dashed lines are guidelines indicating the pressure dependence of transition fields. (b) Enlarged view of the $\Delta f_{\rm sub}$-$H$ curves around $H_{\rm c3}$. The determination of $H_{\rm c3}$ is illustrated. The curves in Figs.~5 (a) and 5 (b) are arbitrarily shifted from the ambient-pressure curve with increasing pressure for clarity.}
\end{figure*}

Figure \ref{fig3}(a) shows the temperature changes from the initial temperature $T_0$ = 1.4 K on the outer wall of the PCC in pulsed magnetic fields as a function of time and the profile of this magnetic fields, which reached a maximum field of  51.0 T, and a duration of 35 ms. The temperature on the outer wall of the PCC rapidly increased as soon as the pulsed magnetic field was generated and exceeded the maximum calibration temperature of 10 K at approximately 20 ms. The thermal equilibrium state between 6 and 15 ms in Fig.\ref{fig3}(a) may be a temporary suppression of the temperature increase owing to the endothermic effect of the evaporation of liquid $^4$He by Joule heating. Figure \ref{fig3}(b) shows the temperature changes from 1.4 K and 4.2 K at the sample position inside the PCC in pulsed magnetic fields as a function of time. At the maximum field of 51.0 T, the temperature at the sample position remained almost 1.4 K until nearly 6.5 ms (approximately 40 T in the field-ascending process). After approximately 6.5 ms, the temperature increased slowly to reach approximately 8 K at 40 ms (approximately zero T). Since the sample is covered with a Teflon tube (the thermal conductivity of Teflon at 2 K is of the order of 10$^{-4}$ (J/cm$\cdot$s$\cdot$K) \cite{reese1965thermal}), and the remaining space is filled with Daphne 7373, the Joule heating from the metal parts of the PCC (the thermal conductivity of NiCrAl at 2 K is of the order of 10$^{-3}$ (J/cm$\cdot$s$\cdot$K)) is transmitted to the sample position with some delays. Therefore, regardless of the maximum magnetic field, the temperature hardly increased until approximately 6.5 ms, after which it increased slowly. At 40 ms, the temperatures at the sample position were 8, 7, and 6 K for $H_{\rm max}$ = 51.0, 41.6, and 27.1 T, respectively. This is because the sweep rate of pulsed magnetic fields ($dH/dt$) increases with increasing the maximum field and the Joule heating becomes large accordingly. At the initial temperature $T_0$ = 4.2 K, the temperature at the sample position gradually increased until about 2.5 ms (approximately 20 T in the field-ascending process), whereupon it increased rapidly. In pulsed magnetic fields of up to 51.0 T, the period of time after which the temperature at the sample position started to increase, was longer at 1.4 K than at 4.2 K. This may be owing to the high thermal conductivity of superfluid helium below 2.17 K that surrounds the PCC immersed in liquid $^4$He.

\section{Study of a triangular-lattice antiferromagnet}

We investigated the magnetic susceptibility of Ba$_3$CoSb$_2$O$_9$, one of the triangular-lattice antiferromagnets (TLAs), using the apparatus developed in this study. The Co$^{2+}$ ions with the effective spin $S$ = 1/2 form an equilateral triangular lattice in the $ab$ plane, indicating both intra- and inter-layer antiferromagnetic exchange interactions \cite{treiber1982ordnungs, doi2004structural}. Below $T_{\rm N}$ = 3.8 K, the magnetic structure at zero field shows a 120$^{\circ}$ spin structure in the $ab$ plane. For $H \parallel ab$ plane, as shown in Fig.\ref{fig4}(a), successive quantum phase transitions occur from the Y coplanar state to the up-up-down (uud) state, and from the uud state to the V state, followed by the V$^{\prime}$ state\cite{susuki2013magnetization}. In this experiment, a plate-shaped single-crystal sample of Ba$_3$CoSb$_2$O$_9$ was placed inside a sensor coil with 25 turns, which was directly wound in the direction perpendicular to the $c$ axis of Ba$_3$CoSb$_2$O$_9$ (inset of Fig.\ref{fig4}(b)). The value of $f_0$ of the PDO was approximately 37 MHz at 4.2 K.

Figure \ref{fig4} (b) shows the changes in the resonance frequencies versus the applied magnetic field ($\Delta f$-$H$) for $H \parallel ab$ plane at 1.4 K and 10 K under ambient pressure without the PCC. The curves of $\Delta f$ vs $H$ exhibit both field ascending and descending processes. The value of $\Delta f$ consists of both the change in the magnetoresistance of the sensor coil and that of coaxial cable in the presence of the magnetic field as the background \cite{Ghannadzadeh}. The $\Delta f$-$H$ curve at 1.4 K indicates distinct frequency shifts corresponding to the changes in the magnetic susceptibility at $H_{\rm c1}$ = 9.4 T, $H_{\rm c2}$ = 15.7 T, $H_{\rm c3}$ = 22.7 T, and $H_{\rm sat}$ = 31.8 T when compared to the $\Delta f$-$H$ curve at 10 K above $T_{\rm N}$.

To obtain the intrinsic magnetic susceptibility of Ba$_3$CoSb$_2$O$_9$, we subtracted the fitting function determined from $\Delta f$ at 10 K, for which the difference from the background data is much greater than that at $T_{\rm N}$ from $\Delta f$ at 1.4 K, and then adjusted the data such that the value of the subtracted $\Delta f_{\rm sub}$ above $H_{\rm sat}$ is constant at zero. The comparison between the $\Delta f_{\rm sub}$-$H$ curve and the field derivative of the magnetization ($dM/dH$) obtained using the conventional induction method is shown in Fig.\ref{fig4}(c). The $\Delta f_{\rm sub}$-$H$ curve agrees very well with $dM/dH$ obtained by the induction method\cite{susuki2013magnetization}. The dip between $H_{\rm c1}$ and $H_{\rm c2}$ corresponds to the uud phase, which exhibits a magnetization plateau at one-third of the saturation magnetization in the magnetization curve. The cusps at $H_{\rm c3}$ and $H_{\rm sat}$ are associated with the magnetic transition from the V to the V$^{\prime}$ phase and the saturation field.

Figure \ref{fig5}(a) demonstrates the $\Delta f_{\rm sub}$-$H$ curves of Ba$_3$CoSb$_2$O$_9$ for $H \parallel ab$ plane at 1.4 K in pulsed magnetic fields of up to 51 T under pressures of up to 1.97 GPa. The $\Delta f_{\rm sub}$-$H$ curve at ambient pressure in the PCC agrees remarkably well with that without the PCC as shown in Figs.~\ref{fig5}(a) and (b), but the noise in the former case exceeds that for the latter. This was probably caused by the poor connection between the sensor coil and the Cu wires passing through the stepped hole of the lower plug. Since pulsed high magnetic fields with the maximum field of 51 T reached approximately 40 T at 6.5 ms from the start of pulsed magnetic field generation, $\Delta f_{\rm sub}$ up to $H_{\rm sat}$ is not affected by the increase in the sample temperature as a result of Joule heating.

With increasing pressure, the peak at $H_{\rm c2}$ shifted to a higher magnetic field, whereas the peaks at $H_{\rm c1}$and $H_{\rm sat}$ stayed almost in place with increasing pressure up to 1.97 GPa. The peak position at $H_{\rm c3}$ does not change against pressure, but the peak at $H_{\rm c3}$ became obscure by the background and was too weak to detect above 1.58 GPa.
Based on the pressure dependence of $H_{\rm sat}$, the intra-layer antiferromagnetic exchange interactions did not change significally. Therefore, the expansion of the uud phase may be accompanied by increasing the effects of thermal and/or quantum fluctuations caused by the relative decrease of the interplanar antiferromagnetic exchange interactions, which enhances the two dimensionality in Ba$_3$CoSb$_2$O$_9$. Another possibility may be a tilting of the sample direction against the magnetic field from the $ab$ plane to the $c$ axis caused by the application of pressure\cite{okada2022field}.

Detailed clarification of the pressure effect on the magnetism in Ba$_3$CoSb$_2$O$_9$ for $H \parallel ab$ plane would require an expansion of the pressure region to beyond 2.1 GPa. The PCC in this study was designed as used in a pulse magnet with a bore diameter of 17$\sim$18 mm. We plan to develop a new PCC with a maximum pressure of 4 GPa by decreasing the inner diameter of the PCC utilized in this study. However, this would shorten the time of heat transfer from the inner wall of pressure cell to the sample position, thus enabling the temperature in the sample space to increase at lower magnetic fields than those in the present study. When we use a pulse magnet with a duration of approximately 200 ms based on our future plan, the magnetic-field sweep rate in the field ascending process would be lowered to approximately 1/5 of that of the pulse magnet used in this study. This long duration might suppress the increase of the sample temperature in the PCC, and thus magnetic susceptibility measurements under higher pressures than 2.1 GPa could be conducted in high magnetic fields.
\section{Summary}

In summary, we developed an apparatus for magnetic-susceptibility-measurements in pulsed magnetic fields of up to 51 T under pressures of up to 2.1 GPa. The temperature at the sample position in our PCC changed slightly until approximately 40 T in the field-ascending process in pulsed high magnetic fields up to the maximum 51 T at 1.4 K. We performed the magnetic susceptibility measurements of the triangular-lattice antiferromagnet Ba$_3$CoSb$_2$O$_9$ in pulsed high magnetic fields under high pressures by the LC method using the PDO technique. We succeeded in observing a change in the resonance frequency that corresponded to the field derivation of the magnetization over the saturation field.

\begin{acknowledgments}
We would like to thank D. Yamamoto for useful discussions. This study was supported by the Sasakawa Scientific Research Grant from the Japan Science Society and JST, the establishment of university fellowships towards the creation of science technology innovation, Grant Number JPMJFS2125. This work was supported by JSPS KAKENHI Grant Numbers JP17H06137, JP17K18758, JP21H01035 and 22K03511.

\end{acknowledgments}

\bibliography{nihongi_rsi}

\end{document}